\documentclass[12pt]{article}

\usepackage[dvips]{graphicx}
\usepackage{epsfig}
\usepackage{amsmath}
\usepackage{amssymb}
\usepackage{float}

\newcommand{\bm}{\begin{multiline}}
\newcommand{\beq}{\begin{equation}}
\newcommand{\eeq}{\end{equation}}
\newcommand{\beqs}{\begin{eqnarray}}
\newcommand{\eeqs}{\end{eqnarray}}
\newcommand{\tr}{\mbox{\rm tr}}
\newcommand{\ra}{\rightarrow}

\begin{document}

\thispagestyle{empty}

\hfill{}

\hfill{}

\hfill{}

\vspace{32pt}

\begin{center}

\textbf{\Large Charged Kaluza-Klein double-black holes in five dimensions }

\vspace{48pt}

{\bf Cristian Stelea,~}
{\bf Kristin Schleich,~}
{\bf and Donald Witt}

\vspace*{0.2cm}

{\it Department of Physics and Astronomy, University of British Columbia}\\
{\it 6224 Agricultural Road, Vancouver, BC V6T 1Z1, Canada}\\[.5em]

\end{center}

\vspace{30pt}

\begin{abstract}
 Using a solution generating technique based on the symmetries of the dimensionally reduced Lagrangian we derive an exact solution of the Einstein-Maxwell-Dilaton field equations in five dimensions describing a system of two general non-extremally charged static Kaluza-Klein black holes with spherical horizons. We investigate some of its thermodynamic properties and we also show how to recover various known solutions in particular cases.
\end{abstract}

\vspace{32pt}

\setcounter{footnote}{0}

\newpage

\section{Introduction}

Exact solutions of Einstein field equations (with or without matter fields) played a key role in the development and the current understanding of gravitational physics in four and higher dimensions. Due to their nonlinearity, Einstein equations cannot be solved in every situation of physical interest and one often has to recourse to simplifying assumptions and make use of various symmetries in order to obtain new exact solutions in a systematic way. Some of the most powerful solution generation techniques in General Relativity have been devised for space-time geometries that belong to the so-called generalized Weyl class, as described in \cite{Emparan:2001wk,Harmark:2004rm}. In $D$-dimensions, solutions in this class assume the existence of $D-2$ abelian Killing vectors. Performing a dimensional reduction on a $(D-3)$-torus and further dualization of the various vector fields to scalars leads to three-dimensional Euclidian gravity coupled to a set of scalar fields. It turns out that, in most cases of interest (as in dimensional reductions of higher-dimensional supergravities), these scalar fields form non-linear sigma-models with coset spaces $G/H$ as the target model \cite{Cremmer:1999du}. Here $G$ is some semi-simple group, $H$ a subgroup of $G$ and the field equations posses then the `hidden symmetry group' $G$, which can be used in various solution generating techniques.

One caveat of this method is that, in order to obtain asymptotically flat black hole solutions, one has to restrict the space-time dimension to $D\leq 5$ since general black holes in $D>5$ dimensions do not admit $(D-2)$-commuting Killing vectors \cite{Emparan:2008eg}. While the four dimensional case has been extensively studied (see for instance \cite{Stephani:tm}), since the recent discovery of black rings \cite{Emparan:2001wn,Emparan:2006mm} there is renewed interest in studying gravity in five and higher dimensions. The black ring was the first explicit example of an asymptotically flat black object with non-spherical event horizon. Heuristically, one obtains such a black ring by taking a black string in five dimensions, bending it and connecting its ends to form a circle. A static black ring configuration would normally collapse  to form a black hole with spherical horizon topology; indeed, this is the case in four dimensional asymptotically flat space-times as a consequence of topological censorship \cite{Friedman:1993ty,Galloway:1999bp}. However, in five or more dimensions, the spherical topology of infinity does not constrain that of the black hole horizon \cite{Galloway:1999br}; geometric considerations, however, restrict the topology to those, such as $S^3$ and $S^2\times S^1$, that admit non-negative scalar curvature \cite{Cai:2001su}. The original black ring system was stabilized by the centrifugal effects of its rotation. Since the black ring discovery, many solution generating techniques were developed to re-derive and further generalize this kind of solutions \cite{Elvang:2003yy,Emparan:2004wy,Yazadjiev:2006hw,Yazadjiev:2006ew,Yazadjiev:2006wr,Iguchi:2006rd,Tomizawa:2006jz,Iguchi:2007is,Tomizawa:2006vp,Koikawa:2005ia,Pomeransky:2005sj,Azuma:2005az,Mishima:2005id,Tomizawa:2005wv,Morisawa:2007di,Elvang:2007rd,Pomeransky:2006bd,Gauntlett:2004wh,Gauntlett:2004qy,Evslin:2007fv,Yazadjiev:2007cd,Elvang:2007hs,Elvang:2007hg, Ida:2003wv,Mei:2007bn}.

In five dimensions there also exist the so-called Kaluza-Klein (KK) black holes. These correspond to black objects whose horizon geometry is a squashed three-sphere \cite{Dobiasch:1981vh,Gibbons:1985ac,Rasheed:1995zv,Larsen:1999pp}. By contrast, their geometry is not asymptotically flat; instead it is asymptotic to a nontrivial $S^1$ bundle with constant fiber over the two-sphere in a four-dimensional asymptotically flat spacetime. This is also the asymptotic geometry of the Kaluza-Klein monopole \cite{Sorkin:1983ns,Gross:1983hb}. Such black holes look five-dimensional in the near horizon region, while asymptotically they look like four-dimensional objects with a compactified fifth dimension.

For vacuum metrics, there exists a systematic procedure to add KK-monopole charge to a general asymptotically flat geometry \cite{Ford:2007th,Giusto:2007fx,Camps:2008hb}, based on a hidden $SL(3,R)$ symmetry of the gravitational sector \cite{Maison:2000fj}.  One should note here that adding a Kaluza-Klein monopole charge to a given solution with matter fields is not a trivial task once one departs from the class of supersymmetric/vacuum solutions;  in many cases, to find exact solutions one has to solve the Einstein equations by brute force. For instance, a solution describing a static KK black hole with electric charge has been found in \cite{Ishihara:2005dp}, while the corresponding Einstein-Yang-Mills solution has been described in \cite{Brihaye:2006ws}. Remarkably, with hindsight, many such KK solutions can be generated by applying a `squashing' transformation on suitable geometries \cite{Wang:2006nw,Nakagawa:2008rm,Tomizawa:2008hw,Matsuno:2008fn,Tomizawa:2008rh,Stelea:2008tt}. However, not all the KK black hole solutions can be generated by the squashing transformation. More recently, in the context of the minimal $5$-dimensional supergravity, there has been developed a solution generating technique based on the $G_2$ U-duality arising in the dimensional reduction of the theory down to three dimensions \cite{Bouchareb:2007ax,Clement:2007qy,Clement:2008qx,Galtsov:2008zz,Gal'tsov:2008nz,Compere:2009zh} and more general KK black holes have been derived \cite{Tomizawa:2008qr,Gal'tsov:2008sh,Chen:2010ih}.

In this article we focus on Kaluza-Klein multi-black hole solutions. In higher dimensions, by contrast to the single black hole case, solutions describing general charged multi-black hole objects are scarce. The main reason is that, except in particular cases where the black objects are extremal \cite{Myers:1986rx,
Duff:1993ye,Ishihara:2006iv,Elvang:2005sa}, the known solution generating techniques lead to multi-black hole systems with charges proportional to the masses and therefore they cannot describe the most general charged solution for which the charges and the masses should be independent parameters. However, in five dimensions, a solution describing a general double-Reissner-Nordstr\"om solution has been recently constructed in \cite{Chng:2008sr}, generalizing the uncharged solutions given in \cite{Tan:2003jz,Teo:2003ug,Teo:2005wf,Kunz:2008rs}. The main purpose of this article is to show that one can appropriately modify the procedure in  \cite{Chng:2008sr} to construct the general charged double KK black hole in five-dimensions.

The structure of this paper is as follows. We first describe the solution generating technique that will allow us to lift four-dimensional charged static configurations to five dimensional Einstein-Maxwell solutions with Kaluza-Klein asymptotics. This technique is based on a simple modification of the procedure given in \cite{Chng:2008sr}. We use the general double Reissner-Nordstr\"om solutions in four dimensions as a seed and lift it to five dimensions and show that the final solution can be interpreted as the general charged Kaluza-Klein double-black hole solution. Our solution generating method extends easily to the more
general case of Einstein-Maxwell-Dilaton (EMD) gravity with arbitrary coupling constant and we derive the charged Kaluza-Klein double-black hole solutions in this case. We end with a summary of our work and consider avenues for future research.

\section{Solution generating technique}

Consider a charged static solution of the Einstein-Maxwell-Dilaton system in five dimensions:
\begin{eqnarray}
\mathcal{L}_{5}=\sqrt{-g}\left[R-\frac{1}{2}(\partial\phi)^2 -\frac{1}{4}%
e^{\alpha\phi}F_{(2)}^2\right],
\label{EMDaction5d}
\end{eqnarray}
where $F_{(2)}=dA_{(1)}$. For static electrically charged configurations the only non-zero component of the $1$-form gauge potential $A_{(1)}$ will be denoted here by $A_{t}$.  The solution is specified by the metric:
\beqs
ds_{(5)}^2&=&-e^{-\frac{2\phi_1}{\sqrt{3}}}dt^2+e^{\frac{\phi_1}{\sqrt{3}}}\big[e^{-\psi}(d\chi-\omega d\varphi)^2+e^{\psi}ds_{(3)}^2\big],
\label{KKansatz5d}
\eeqs
the scalar field $\phi$ and the gauge field $A_{(1)}=A_tdt$.
Here and in what follows we assume that all the functions $\phi_1$, $\psi$, $\omega$, $A_{t}$ and the
scalar field $\phi$ depend only on coordinates $\rho$ and $z$.

Perform now a dimensional reduction down to three dimensions: \textit{first} on the time direction and then on the $\chi$ coordinate\footnote{The order is important here since if one performs a KK reduction first on $\chi$ and then on $t$ one gets `mixed' terms in the three-dimensional Lagrangian and further field redefinitions of the scalar fields have to be made to decouple the ($\phi_1$, $A_t$, $\phi$) and ($\psi$, $\omega$) sectors.}. Denoting the KK $1$-form potential from the metric by ${\cal A}_{(1)}=\omega d\varphi$ one obtains a solution of the following Lagrangian in three dimensions:
\beqs
{\cal L}_{(3)}&=&\sqrt{g}\bigg[R-\frac{1}{2}(\partial\phi)^2-\frac{1}{2}(\partial\phi_1)^2+\frac{1}{2}e^{\frac{2}{\sqrt{3}}\phi_1+\alpha \phi}(\partial A_t)^2-\frac{1}{2}(\partial\psi)^2-\frac{1}{4}e^{-2\psi}(d{\cal A}_{(1)})^2\bigg].
\label{L3}
\eeqs
One should note at this point that one has two decoupled systems: one involving the scalar fields $\phi$, $\phi_1$ and the electric potential $A_t$, while the second sector comprises the scalar $\psi$ and the KK $1$-form ${\cal A}_{(1)}$.  This simple observation is the base of our solution generation technique in the present work. To this end, we shall employ Weyl's charging technique as described for instance in \cite{Chng:2006gh}. Let us first perform a dualization of the $2$-form field strength ${\cal F}_{(2)}=d{\cal A}_{(1)}$ to a scalar $\xi$ by using:
\beqs
{\cal F}_{ab}&=&\sqrt{g}e^{2\psi}\epsilon_{abc}\partial^c\xi,
\label{dualF}
\eeqs
where $\epsilon_{abc}$ is the Levi-Civita symbol in three dimensions. Then one can rewrite the $(\psi,{\cal A}_{(1)})$ sector in the above Lagrangian in the following form:
\beqs
{\cal L}_{(\psi,{\cal A})}&=&\sqrt{g}\bigg[ -\frac{1}{2}(\partial\psi)^2+\frac{1}{2}e^{2\psi}(\partial\xi)^2\bigg]=\sqrt{g} \frac{1}{4} \tr\big[\partial{\cal M}^{-1}\partial{\cal M}\big],
\label{Lpsi}
\eeqs
where one defines the matrix:
\beqs
\cal{M}=\left(\begin{array}{cc}  e^{\psi} &\xi e^{\psi} \\
 \xi e^{\psi} & -e^{-\psi}+\xi^2e^{\psi} \end{array}\right).
\eeqs
It is now manifest that the truncated Lagrangian (\ref{Lpsi}) is invariant under general $SL(2,R)$ transformations if one considers the following transformation law:
\beqs
{\cal M}\ra\Omega^T{\cal M}\Omega,~~~~~~~\Omega=\left(\begin{array}{cc}a&b\\ c&d\end{array}\right), ~~~~ad-bc=1.
\eeqs
The scalar fields $\psi$ and $\xi$ transform under $\Omega$ as follows:
\beqs
e^{\psi'}&=&a^2e^{\psi}+2ac\xi e^{\psi}-c^2e^{-\psi}+c^2\xi^2e^{\psi},\nonumber\\
\xi'e^{\psi'}&=&abe^{\psi}+(ad+bc)\xi e^{\psi}-dce^{-\psi}+dc\xi^2e^{\psi}.
\eeqs
Suppose now that one starts with a solution (\ref{KKansatz5d}) for which $\omega=0$, that is $\xi=0$. Then in terms of the initial scalar (denote it by $\tilde{\psi}$) after applying an $SL(2,R)$ transformation (the metric and the remaining fields in (\ref{L3}) remain unchanged) one obtains:\footnote{For further convenience one drops the prime superscripts in the final fields.}
\beqs
e^{\psi}&=&e^{\tilde{\psi}}(a^2-c^2e^{-2\tilde{\psi}}),~~~~~~~
\xi=-\frac{c}{a(a^2e^{2\tilde{\psi}}-c^2)},
\eeqs
up to the constant term $b/a$ in $\xi$ that can be dropped without losing generality. Next, to obtain a solution of the system (\ref{L3}) one has to dualize back the scalar field $\xi$ to obtain the KK $1$-form ${\cal A}_{(1)}=\omega d\varphi$. Replacing the above expression of $\xi$ in (\ref{dualF}) it turns out that the problem reduces to find a function $\omega=2ac\Psi$ such that:
\beqs
d\Psi&=&\rho(\partial_{\rho}\tilde{\psi} dz-\partial_z\tilde{\psi} d\rho)
\eeqs
The integrability condition for $\Psi$ leads to following condition on $\tilde{\psi}$:
\beqs
\partial_{\rho}^2\tilde{\psi}+\frac{1}{\rho}\partial_{\rho}\tilde{\psi}+\partial_z^2\tilde{\psi}=0,
\eeqs
that is $\tilde{\psi}$ must be a harmonic function. This condition is automatically satisfied for any initial seed solution for which $\omega=0$ and, therefore, it does not impose any real constraint. As in \cite{Giusto:2007fx}, given a harmonic function $\tilde{\psi}$ we shall call a function $\Psi$ satisfying the above equation as `the dual' of $\tilde{\psi}$ and it turns out that, when $\tilde{\psi}$ is given in terms of simple rods, one can easily write down the expression for $\omega=2ac\Psi$ in closed form. Once $\omega$ and $\psi$ are known one can replace them in (\ref{KKansatz5d}) to obtain the new charged solution of the EMD system in five dimensions, all the remaining fields being unchanged under the action of the $SL(2,R)$ transformation.

In what follows we shall apply this technique on the charged static five-dimensional solutions obtained in  \cite{Chng:2008sr}. Let us recall first the results of the solution generating technique used in that work. Suppose we are given a static solution of the four-dimensional Einstein-Maxwell system:
\begin{eqnarray}  \label{4delectric}
\mathcal{L}_4&=&\sqrt{-g}\left[R-\frac{1}{4}\tilde{F}_{(2)}^2\right],
\end{eqnarray}
where $\tilde{F}_{(2)}=d\tilde{A}_{(1)}$ and the only non-zero component of $%
\tilde{A}_{(1)}$ is $\tilde{A}_{t}=\Phi$. The solution to the equations of
motion derived from (\ref{4delectric}) is assumed to have the following
static and axisymmetric form:
\begin{eqnarray}  \label{newinitialmetric1}
ds_{4}^{2} &=&-\tilde{f}dt^{2}+\tilde{f}^{-1}\big[e^{2\tilde{\mu}}(d\rho
^{2}+dz^{2})+\rho ^{2}d\varphi ^{2}\big],  \notag \\
{\tilde{A}_{(1)}} &=&\Phi dt.
\end{eqnarray}

Then the corresponding solution of the Einstein-Maxwell-Dilaton system in five dimensions can be written as:
\begin{eqnarray}  \label{new5Dkfl}
ds_{5}^{2}=-\tilde{f}^{\frac{4}{3\alpha^2+4}}dt^{2}+\tilde{f}^{-\frac{2}{%
3\alpha^2+4}}\bigg[e^{2h}d\chi ^{2}+e^{\frac{6\tilde{\mu}}{3\alpha^2+4}%
+2\gamma-2h}(d\rho ^{2}+dz^{2})+\rho^2e^{-2h}d\varphi ^{2}\bigg],
\end{eqnarray}
while the $1$-form potential and the dilaton are given by:
\begin{eqnarray}
A_{(1)}&=&\sqrt{\frac{3}{3\alpha^2+4}}\Phi dt,~~~~~~~ e^{-\phi}=\tilde{f}^{%
\frac{3\alpha}{3\alpha^2+4}}.
\end{eqnarray}
It can be checked that this solution solves the equations of motion derived from (\ref{EMDaction5d}).
Here $h$ is an arbitrary harmonic function; once its form has been specified for a particular solution then $\gamma$ can be obtained by simple quadratures using the equations:
\begin{eqnarray}  \label{gammap1a}
\partial_\rho{\gamma}&=&\rho[(\partial_\rho h)^2-(\partial_z h)^2],~~~~~~~
\partial_z{\gamma}=2\rho(\partial_\rho h)(\partial_z h).
\end{eqnarray}
It was shown in \cite{Chng:2008sr} that, using the four-dimensional Reissner-Nordstr\"om as the initial seed, by an appropriate choice of $h$ one can obtain either a black hole, or a black ring or a black string in five dimensions.

Recasting (\ref{new5Dkfl}) in the form (\ref{KKansatz5d}) one can read directly the following fields:
\beqs
e^{\frac{\phi_1}{\sqrt{3}}}&=&\tilde{f}^{-\frac{2}{3\alpha^2+4}},~~~~~\tilde{\psi}=-2h,~~~~~~~ds_{(3)}^2=e^{\frac{6\tilde{\mu}}{3\alpha^2+4}+2\gamma}(d\rho ^{2}+dz^{2})+\rho^2d\varphi ^{2}.
\eeqs
If one denotes by $H$ the `dual' of $h$ then the final solution of the EMD system can be expressed as:
\beqs
\label{final5dalpha}
ds_{5}^{2}&=&-\tilde{f}^{\frac{4}{3\alpha^2+4}}dt^{2}+\tilde{f}^{-\frac{2}{3\alpha^2+4}}\bigg[\frac{e^{2h}}{a^2-c^2e^{4h}}(d\chi+4acH d\varphi)^{2}+(a^2-c^2e^{4h})e^{\frac{6\tilde{\mu}}{3\alpha^2+4}
+2\gamma-2h}(d\rho ^{2}+dz^{2})\nonumber\\
&&+\rho^2(a^2-c^2e^{4h})e^{-2h}d\varphi ^{2})\bigg],~~~~~~~
A_{(1)}=\sqrt{\frac{3}{3\alpha^2+4}}\Phi dt,~~~~~~~ e^{-\phi}=\tilde{f}^{\frac{3\alpha}{3\alpha^2+4}}.
\eeqs

Solutions of the pure Einstein-Maxwell theory in five dimensions are simply obtained from the above formulae by taking $\alpha=0$. In the following sections we shall focus on this particular case.

\section{Charged Kaluza-Klein black hole solutions}

As a check of the above solution generating technique we shall first generate the single charged KK black hole, which was originally derived in \cite{Ishihara:2005dp}. Not too much of a surprise, it turns out that in order to generate this solution one has to employ the charged black string as the initial seed. Motivated by this result, we then use the double-black string solution constructed in \cite{Chng:2008sr} to derive the general charged static KK double-black hole system in five dimensions.

\subsection{Single charged KK black hole}

Let us start with the five-dimensional black string solution. In Weyl coordinates it can be written in the following form \cite{Chng:2008sr}:
\beqs
ds_{(5)}^2&=&-\tilde{f}dt^2+\tilde{f}^{-\frac{1}{2}}\bigg[e^{2h}d\chi^2+e^{-2h}\big[e^{\frac{3\tilde{\mu}}{2}+2\gamma}(d\rho^2+dz^2)+\rho^2d\varphi^2\big]\bigg],
\eeqs
where:
\beqs
\tilde{f}&=&\frac{(r_1+r_2)^2-4\sigma^2}{(r_1+r_2+2m)^2},~~~~~e^{2\tilde{\mu}}=\frac{(r_1+r_2)^2-4\sigma^2}{4r_1r_2},~~~~~A_{t}=-\frac{2\sqrt{3}q}{r_1+r_2+2m},\nonumber\\
e^{2h}&=&\sqrt{\frac{r_1+\zeta_1}{r_2+\zeta_2}},~~~~~e^{2\gamma}=\frac{1}{K_0}\left(\frac{16Y_{12}}{r_1r_2}\right)^{\frac{1}{4}}.
\eeqs
Here we denote by $\sigma=\sqrt{m^2-q^2}$, $r_i=\sqrt{\rho^2+\zeta_i^2}$, $\zeta_1=z-\sigma$, $\zeta_2=z+\sigma$, while $2Y_{12}=(r_1+r_2)^2-4\sigma^2$. Using the explicit form for $h$ given above, it is straightforward to verify that its dual is given by $H=\frac{r_2-r_1}{4}$.\footnote{In general, the dual of $\frac{1}{2}\ln(r_i+\zeta_i)$ is $-\frac{1}{2}(r_i-\zeta_i)$, while the dual of $\frac{1}{2}\ln(r_i-\zeta_i)$ is $-\frac{1}{2}(r_i+\zeta_i)$, where $r_i=\sqrt{\rho^2+\zeta_i^2}$, $\zeta_i=z-a_i$ and $a_i$ is constant.}

The final solution can be written in the form:
\beqs
\label{singleKK5d}
ds_{(5)}^2&=&-\frac{(r_1+r_2)^2-4\sigma^2}{(r_1+r_2+2m)^2}dt^2+\frac{r_1+r_2+2m}{\sqrt{2Y_{12}}}\bigg[\frac{r_2+\zeta_2}{a^2(r_2+\zeta_2)-c^2(r_1+\zeta_1)}\sqrt{\frac{r_1+\zeta_1}{r_2+\zeta_2}}(d\chi+ac(r_2-r_1)d\varphi)^2\nonumber\\
&&+\frac{a^2(r_2+\zeta_2)-c^2(r_1+\zeta_1)}{r_2+\zeta_2}\sqrt{\frac{r_2+\zeta_2}{r_1+\zeta_1}}\left(\frac{2^{\frac{1}{4}}Y_{12}}{K_0r_1r_2}(d\rho^2+dz^2)+\rho^2d\varphi^2\right)\bigg].
\eeqs

Before we show that this is indeed the charged KK black hole solution, let us notice that for $q=0$, $a=1$, $\sigma=k^2$ and $c=\frac{2l^2-\sigma}{2l^2+\sigma}$ it reduces to the uncharged KK black hole derived in \cite{Ford:2007th}. Perform now the following coordinate changes:
\beqs
\rho&=&\sqrt{(r-m)^2-\sigma^2}\sin\theta,~~~~~~~z=(r-m)\cos\theta.
\eeqs
Noting that $r_2-r_1=2\sigma\cos\theta$ and picking $K_0=2^{\frac{5}{4}}$ it is now apparent that (\ref{singleKK5d}) becomes:
\beqs
ds_{(5)}^2&=&-\frac{(r-m)^2-\sigma^2}{r^2}dt^2+g(r)\bigg[\frac{dr^2}{\frac{(r-m)^2-\sigma^2}{r^2}}+r^2(d\theta^2+\sin^2\theta d\varphi^2)\bigg]+\frac{1}{g(r)}(d\chi+2ac\sigma\cos\theta d\varphi)^2,\nonumber\\
A_t&=&-\frac{\sqrt{3}\sqrt{m^2-\sigma^2}}{r},~~~~~~~g(r)=a^2-c^2+\frac{\sigma(a^2+c^2)-m(a^2-c^2)}{r}.
\label{first5dsingle}
\eeqs
This is none other than the charged static Kaluza-Klein black hole derived in \cite{Ishihara:2005dp}. To see this, let us fix $a$ such that $a^2=c^2+1$ and redefine the constants in our solution such that $r_{\infty}=2c\sqrt{c^2+1}\sigma$ and $r_0=\sigma(1+2c^2)-m$. By expressing $c$ in terms of $r_{\infty}$ and $\sigma$ it can be easily checked that (\ref{first5dsingle}) becomes:
\beqs
ds^2&=&-f(r)dt^2+g(r)\left(\frac{dr^2}{f(r)}+r^2(d\theta^2+\sin^2\theta d\varphi^2)\right)+\frac{r_{\infty}^2}{g(r)}(d\chi+\cos\theta d\varphi)^2,\nonumber\\
f(r)&=&\frac{(r-m)^2-\sigma^2}{r^2},~~~g(r)=1+\frac{r_0}{r}=1+\frac{\sqrt{r_{\infty}^2+\sigma^2}-m}{r},
\eeqs
that is, the solution found in \cite{Ishihara:2005dp}.

The extremal limit is achieved when $\sigma=0$ while the uncharged black hole is obtained by taking $\sigma=m$. In absence of the black hole (take $m=\sigma=0$) the solution becomes the KK-monopole background as expected. This solution is described in terms of the length at infinity, $4\pi r_{\infty}$, of the $\chi$ coordinate.

In addition, it can be manifestly checked that the Reissner-Nordstr\"om black hole with spherical  horizon is recovered from (\ref{first5dsingle}) if one takes directly $a=c$ and makes the analytical continuation $c\rightarrow ic$ to keep $g(r)$ positive.

One could also apply the above solution generating technique using as seed either the five-dimensional Reissner-Nordstr\"om black hole with spherical horizon or the static charged black ring. However, the generated solutions do not seem to have the right KK asymptotics and, therefore, they are not amenable to clear physical interpretations. For instance, if one starts with the spherical black hole, after converting the final solution from the Weyl coordinates to the spherical ones, in the uncharged case one obtains the following metric:
\beqs
ds_{(5)}^2&=&-\left(1-\frac{4m}{r^2}\right)dt^2+(a^2-c^2r^2(r^2-4m)\cos^4\theta)\left(\frac{dr^2}{1-\frac{4m}{r^2}}+r^2(d\theta^2+\sin^2\theta d\varphi^2)\right)\nonumber\\
&&+\frac{r^2\cos^2\theta}{a^2-c^2r^2(r^2-4m)\cos^4\theta}(d\chi-2ac(r^2-2m)\sin^2\theta d\varphi)^2,
\eeqs
upon performing the coordinate transformations:
\beqs
\rho^2&=&r^2(r^2-4m)\sin^2\theta\cos^2\theta,~~~~~~~z=\frac{1}{2}(r^2-2m)\cos2\theta.
\eeqs
We explicitly checked that this is indeed a vacuum solution, however it has peculiar asymptotics and its physical interpretation (if any) is unclear to us at this stage.

\subsection{Charged Kaluza-Klein double-black hole solution}

As we have seen in the previous section, in order to generate the charged KK black hole one has to use the charged black string as seed. To obtain the general charged KK double-black hole solution one should then use as initial seed in our solution generating technique the charged double-black string solution constructed in \cite{Chng:2008sr}. In Weyl coordinates  this solution can be written as:
\begin{eqnarray}  \label{rel1}
ds_{5}^2&=&-\tilde{f}dt^2+\tilde{f}^{-\frac{1}{2}}\bigg[e^{2h}d\chi^2+e^{-2h}%
\big[e^{3\tilde{\mu}/2+2\gamma}(d\rho^2+dz^2)+\rho^2d\varphi^2\big]\bigg],
\notag \\
A_t&=&-\frac{\sqrt{3}C}{A+B}.  \label{final5}
\end{eqnarray}
Here one denotes (see \cite{Manko:2007hi} for details of the original double Reissner-Nordstr\"om solution in four dimensions):
\begin{equation}
\tilde{f}=\frac{A^{2}-B^{2}+C^{2}}{(A+B)^{2}},~~~~~~e^{2\tilde{\mu}}=\frac{%
A^{2}-B^{2}+C^{2}}{16\sigma _{1}^{2}\sigma _{2}^{2}(\nu
+2k)^{2}r_{1}r_{2}r_{3}r_{4}},  \label{Manko}
\end{equation}%
where:
\begin{eqnarray}
A &=&\sigma _{1}\sigma _{2}[\nu
(r_{1}+r_{2})(r_{3}+r_{4})+4k(r_{1}r_{2}+r_{3}r_{4})]-(\mu ^{2}\nu
-2k^{2})(r_{1}-r_{2})(r_{3}-r_{4}),  \notag \\
B &=&2\sigma _{1}\sigma _{2}[(\nu M_{1}+2kM_{2})(r_{1}+r_{2})+(\nu
M_{2}+2kM_{1})(r_{3}+r_{4})]  \notag \\
&&-2\sigma _{1}[\nu \mu (Q_{2}+\mu )+2k(RM_{2}+\mu Q_{1}-\mu
^{2})](r_{1}-r_{2})  \notag \\
&&-2\sigma _{2}[\nu \mu (Q_{1}-\mu )-2k(RM_{1}-\mu Q_{2}-\mu
^{2})](r_{3}-r_{4}),  \notag \\
C &=&2\sigma _{1}\sigma _{2}\{[\nu (Q_{1}-\mu )+2k(Q_{2}+\mu
)](r_{1}+r_{2})+[\nu (Q_{2}+\mu )+2k(Q_{1}-\mu )](r_{3}+r_{4})\}  \notag \\
&&-2\sigma _{1}[\mu \nu M_{2}+2k(\mu M_{1}+RQ_{2}+\mu R)](r_{1}-r_{2})
\notag \\
&&-2\sigma _{2}[\mu \nu M_{1}+2k(\mu M_{2}-RQ_{1}+\mu R)](r_{3}-r_{4}),
\end{eqnarray}%
with constants:
\begin{eqnarray}
\nu &=&R^{2}-\sigma _{1}^{2}-\sigma _{2}^{2}+2\mu
^{2},~~~~~~~k=M_{1}M_{2}-(Q_{1}-\mu )(Q_{2}+\mu ),  \notag \\
\sigma _{1}^{2} &=&M_{1}^{2}-Q_{1}^{2}+2\mu Q_{1},~~~~~~~\sigma
_{2}^{2}=M_{2}^{2}-Q_{2}^{2}-2\mu Q_{2},~~~~~~~\mu =\frac{%
M_{2}Q_{1}-M_{1}Q_{2}}{M_{1}+M_{2}+R},
\end{eqnarray}%
while $r_{i}=\sqrt{\rho ^{2}+\zeta _{i}^{2}}$, for $i=1..4$, with:
\begin{equation}
\zeta _{1}=z-\frac{R}{2}-\sigma _{2},~~~~~\zeta _{2}=z-\frac{R}{2}+\sigma
_{2},~~~~~\zeta _{3}=z+\frac{R}{2}-\sigma _{1},~~~~~\zeta _{4}=z+\frac{R}{2}%
+\sigma _{1}.
\label{ai}
\end{equation}%

 To describe a configuration of two charged black strings one has to pick the following harmonic function $h$, while $\gamma$  can be found by integrating (\ref{gammap1a}) with the result:
 \begin{eqnarray}
e^{2h}&=&\sqrt{\frac{(r_1+\zeta_1)(r_3+\zeta_3)}{(r_2+\zeta_2)(r_4+\zeta_4)}},~~~~~~~
e^{2\gamma}=\frac{1}{K_0}\left(\frac{16Y_{12}Y_{14}Y_{23}Y_{34}}{%
r_1r_2r_3r_4Y_{13}Y_{24}}\right)^{\frac{1}{4}},
\end{eqnarray}
where $K_0$ is a constant and we denote $Y_{ij}=r_ir_j+\zeta_i\zeta_j+\rho^2$, where $i,j=1...4$. Let us note that the dual of $h$ is given by:
\beqs
H&=&\frac{r_2+r_4-r_1-r_3}{4},
\eeqs
while
\beqs
\Sigma\equiv a^2-c^2e^{4h}&=&\frac{a^2(r_2+\zeta_2)(r_4+\zeta_4)-c^2(r_1+\zeta_1)(r_3+\zeta_3)}{(r_2+\zeta_2)(r_4+\zeta_4)}.
\eeqs
The final solution describing a system of two general charged KK black holes in five dimensions is given by:
\beqs
ds_{(5)}^2&=&-\tilde{f}dt^2+\tilde{f}^{-\frac{1}{2}}\bigg[\frac{e^{2h}}{\Sigma}(d\chi+ac(r_2+r_4-r_1-r_3)d\varphi)^2+e^{-2h}\Sigma
\big[e^{3\tilde{\mu}/2+2\gamma}(d\rho^2+dz^2)+\rho^2d\varphi^2\big]\bigg],
\notag \\
A_t&=&-\frac{\sqrt{3}C}{A+B}.  \label{final5}
\end{eqnarray}
This solution is parameterized by six independent parameters and describes the superposition of two general KK black holes, with masses $M_{1,2}$, charges $Q_{1,2}$, $R$ being the coordinate distance
separating them and the asymptotic length of the KK coordinate $\chi$ at infinity, or, equivalently, by the KK-monopole charge.

Before we discuss some of its physical properties, let us first consider the rod structure of this solution. Following the procedure given in \cite{Harmark:2004rm} one deduces that the rod structure of the general solution is described by four turning points that divide the $z$-axis into five rods as follows.\footnote{We are writing the vectors in the basis $\{\partial/\partial t, \partial/\partial \chi,\partial/\partial\varphi\}$.} For $z<-R/2-\sigma_1$ one has a semi-infinite rod with direction $l_1=(0,2ac(\sigma_1+\sigma_2),1)$, for $-R/2-\sigma_1<z<-R/2+\sigma_1$ one has a timelike rod with direction $l_2=(1,0,0)$, which corresponds to the horizon of the first black hole. For $-R/2+\sigma_1<z<R/2-\sigma_2$ on has a finite spacelike rod with direction $l_3=(0,-2ac(\sigma_1-\sigma_2),1)$. For $R/2-\sigma_2<z<R/2+\sigma_2$ one has a timelike rod with direction $l_4=(1,0,0)$, which corresponds to the horizon of the second black hole. Finally, for $z>R/2+\sigma_2$ one has a semi-infinite spacelike rod with direction $l_5=(0,-2ac(\sigma_1+\sigma_2),1)$. Note now that the rod directions of the spacelike rods surrounding the horizons are precisely the rod directions of the double-Taub-NUT background \cite{Chen:2010zu}. This confirms that the general solution that we derived describes a pair of black holes sitting at the nuts of the double Taub-NUT background. 

 Turning now to the discussion of the conical singularities, to avoid a conical singularity at the location of a rod with direction $l_i$, the period $\Delta_i$ of the spacelike coordinate $\eta_i$ (such that $l_i=\partial/\partial \eta_i$) must be fixed as:
 \beqs
 \Delta_i=2\pi\lim_{\rho\ra 0}\sqrt{\frac{\rho^2g_{\rho\rho}}{|l_i|^2}},
 \eeqs
 where $g_{\rho\rho}$ is the $\rho\rho$-component of the metric while $|l_i|^2$ is the norm of $l_i$. Specifically we find:
\begin{eqnarray}
\Delta_{1}&=&\Delta_5=2\pi\sqrt{\frac{\sqrt{8}}{K_0}}
\end{eqnarray}
along the outer axis $z<-R/2-\sigma_1$ or $z>R/2+\sigma_2$, while:
\begin{eqnarray}
\Delta_3&=&2\pi\sqrt{\frac{\sqrt{8}}{K_0}}\left(\left(\frac{\nu-2k}{\nu+2k}\right)^{3}\frac{R^2-(\sigma_1-\sigma_2)^2}{R^2-(\sigma_1+\sigma_2)^2}\right)^{\frac{1}{4}},
\end{eqnarray}
on the portion $-R/2+\sigma_1<z<R/2-\sigma_2$ in between the KK black hole
horizons. We ensure regularity of the outer axis, by taking $K_0=\sqrt{8}$. There will still be a conical singularity running in between the KK black holes. The equilibrium
condition, for which this conical singularity disappears is given by:
\begin{eqnarray}  \label{conditie-UBS}
\left(\frac{\nu-2k}{\nu+2k}\right)^3&=&\frac{R^2-(\sigma_1+\sigma_2)^2}{R^2-(\sigma_1-\sigma_2)^2}.
\end{eqnarray}

This is the same equilibrium condition found in \cite{Chng:2006gh} when discussing the double black string solution in the KK flat background. Therefore, as in \cite{Chng:2006gh} we conclude that there are no nonextremal KK double-black hole solutions with $\delta_{\varphi}=0$, which also satisfy the physical conditions $M_1+M_2>0$ and $\sigma_1+\sigma_2<R$. The only way to satisfy this condition is to consider extremal objects for which $\sigma_1=\sigma_2=0$.

\subsubsection{Limits of the KK double-black hole solution}

Let us consider now some particular cases of the above general charged KK double-black hole solution. First, in order to prove that it describes indeed a system of two charged KK static black holes, let us note that one can recover the individual black holes by pushing the other black hole to infinity. For example, to recover the metric describing the second black hole (with parameters $M_2$ and $Q_2$) one has to first center the $z$ origin on its horizon by making the shift $z\rightarrow z-\frac{R}{2}$ and, after that, take the infinite separation limit $R\rightarrow\infty$. From the general expressions in (\ref{Manko}), one notes that in this limit $\nu\sim R^2$, $\mu\sim 0$, $\sigma_i=\sqrt{M_i^2-Q_i^2}$, for $i=1..2$, $k=M_1M_2-Q_1Q_2$ and $r_{3,4}\sim R$, such that:
\begin{eqnarray}
A\sim2\sigma_1\sigma_2R^3(r_1+r_2),~~~~~B\sim4\sigma_1\sigma_2R^3M_2,~~~~~C\sim4\sigma_1\sigma_2R^3Q_2,
\label{pushaway}
\end{eqnarray}
Also, by taking this limit in the harmonic function $h$ one obtains:
\begin{eqnarray}
e^{2h}&=&\sqrt{\frac{r_{1}+\zeta_1}{r_{2}+\zeta_2}},~~~~~~~ e^{2\gamma}=%
\frac{1}{K_0}\left(\frac{16Y_{12}} {r_{1}r_{2}}\right)^{\frac{1}{4}}.
\end{eqnarray}
while the dual of $h$ and $\Sigma$ become respectively:
\beqs
H&=&-\frac{r_1-r_2}{4},~~~~~~~\Sigma^{-1}=\frac{r_2+\zeta_2}{a^2(r_2+\zeta_2)-c^2(r_1+\zeta_1)}.
\label{extremal1limit}
\eeqs
 Replacing all these quantities in (\ref{final5}) it is now manifest that one recovers the solution describing a single KK black hole. However, since in this case $K_0=\sqrt{8}$ there is still a conical singularity along the $\varphi$-axis all the way to infinity as expected.

The extremally charged limit formally corresponds to taking $M_1=\epsilon Q_1$ and $M_2=\epsilon Q_2$ with $\epsilon=\pm 1$. This leads to $\sigma_1=\sigma_2=k=\mu=0$ and, in consequence, in this limit $r_1=r_2$ while $r_3=r_4$. Apparently, this makes $e^{2h}\rightarrow1$ and $H\rightarrow0$. Therefore, in this unconstrained limit the general solution given in (\ref{final5}) reduces to the extremal double black string seed from \cite{Chng:2008sr}, which describes a pair of extremal strings in a flat background. In order to recover the solution describing extremal black holes a background with KK asymptotics, the extremal limit must be taken with better care, as we shall describe in what follows.  

To this end, let us first set $a^2=1+c^2$ and, for future convenience, perform a shift $z\ra z+R/2$ of the $z$ coordinate such that one centers on the horizon of the second black hole.  In the extremal limit one has $\sigma_1,\sigma_2\ra 0$. To correctly obtain this limit one can consider for instance values of the masses or charges such that $\sigma_1=N_1/2c^2$ and $\sigma_2=N_2/2c^2$, where $N_1$ and $N_2$ are constants. Then the extremal limit is simply obtained by taking $c\ra\infty$ such that the products $2c^2\sigma_1\equiv N_1$ and $2c^2\sigma_2\equiv N_2$ are kept constant. One can check that in this constrainted limit one also obtains $M_1=\epsilon Q_1$ and $M_2=\epsilon Q_2$ with $\epsilon=\pm 1$ as expected. Expanding to first order in $\sigma_1$ and $\sigma_2$ one obtains\footnote{Note that only the terms linear in $\sigma_i$ will survive in the extremal limit with $N_i$ kept constants, while terms with higher powers of $\sigma_i$ will vanish.}:
\beqs
\Sigma=1+\frac{N_2}{\sqrt{\rho^2+z^2}}+\frac{N_1}{\sqrt{\rho^2+(z+R)^2}}+{\cal O}(\sigma_i\sigma_j),
\eeqs
while
\beqs
ac(r_2-r_1+r_4-r_3)=\frac{N_2z}{\sqrt{\rho^2+z^2}}+\frac{N_1(z+R)}{\sqrt{\rho^2+(z+R)^2}}+{\cal O}(\sigma_i\sigma_j).
\eeqs

On the other hand, once we have $\sigma_1=\sigma_2=0$ then (\ref{Manko}) becomes:
\begin{eqnarray}
\tilde{f}_e&=&\left(1+\frac{M_1}{r_3}+\frac{M_2}{r_1}\right)^{-2},~~~~~~~
e^{2\tilde{\mu}}|_{e}=1.  \label{extremal4dManko}
\end{eqnarray}
while\footnote{Recall that we set $K_0=\sqrt{8}$ to cancel out conical singularities on the outer axis, leaving a conical singularity on the portion of the $\varphi$-axis in between the black holes.}
\beqs
e^{2h}&=&1,~~~e^{2\gamma}=1.
\eeqs
Collecting all these results into (\ref{final5}) and performing the coordinate transformations:
\beqs
\rho&=&r\sin\theta,~~~~~~~z=r\cos\theta
\eeqs
the extremal double black hole solution becomes:
\beqs
\label{extremalemparan}
ds_{e}^2&=&-\left(1+\frac{M_1}{r_3}+\frac{M_2}{r}\right)^{-2}dt^2+\left(1+%
\frac{M_1}{r_3}+\frac{M_2}{r}\right)\bigg[\frac{1}{1+\frac{N_2}{r}+\frac{N_1}{r_3}}\left(d\chi+\omega_{\varphi} d\varphi\right)^2\nonumber\\&&+\left(1+\frac{N_2}{r}+\frac{N_1}{r_3}\right)\big[dr^2+r^2(d\theta^2+\sin^2\theta d\varphi^2)\big]\bigg],  \nonumber \\
A_t&=&-\frac{\sqrt{3}}{2}\left(1+\frac{M_1}{r_3}+\frac{M_2}{r}\right)^{-1},
\eeqs
where $r_3=\sqrt{r^2+2Rr\cos\theta+R^2}$, while
\beqs
\omega_{\varphi}=N_2\cos\theta+\frac{N_1(r\cos\theta+R)}{\sqrt{r^2+2rR\cos\theta+R^2}}.
\eeqs

 It is interesting to note that the above solution is a particular case of the extremal solutions considered in \cite{Ishihara:2006iv} as expected\footnote{We thank the anonymous referee for stressing this point to us.}. It corresponds to a pair of extremal Kaluza-Klein black holes on a double-Taub-NUT background. In absence of the black holes (setting $M_{1,2}=0$), the solution reduces to the double Taub-NUT background. To summarize, we found that the unconstrained limit yields 2 extremal black string solutions but the alternate limit, in which one takes the limit with the constraints that $ 2c^2 \sigma_1\equiv N_1$ and $2 c^2 \sigma_2 \equiv N_2$ are fixed yields another, the extremal double black hole solution.

\subsubsection{Conserved charges and thermodynamics}

To find the asymptotic geometry one performs the coordinate transformations:
\beqs
\rho&=&r\sin\theta,~~~~~~~z=r\cos\theta,
\eeqs
and take the asymptotic limit $r\ra\infty$. Defining now $r_{\infty}=2c\sqrt{1+c^2}(\sigma_1+\sigma_2)$ the asymptotic length of the $\chi$-circle becomes ${\cal L}=4\pi r_\infty$.

To compute the conserved charges we shall make use of the counterterm proposed in \cite{Mann:2005cx}. This counterterm regularizes the gravitational action for spacetime geometries that are asymptotic to the KK-monopole background. To this end, one adds to the Einstein-Maxwell action (including the Gibbons-Hawking term) the following surface term:
\begin{equation}
I_{ct}=\frac{1}{8\pi G}\int d^{4}x\sqrt{-h}\sqrt{2\mathcal{R}}  \label{I2}
\end{equation}%
where $\mathcal{R}$ is the Ricci scalar of the induced metric on the
boundary, $h_{ij}$. By taking the variation of this total action with
respect to the boundary metric $h_{ij}$, it is straightforward to compute
the boundary stress-tensor:
\begin{equation*}
T_{ij}=\frac{1}{8\pi G}\left( K_{ij}-Kh_{ij}-\Psi( \mathcal{R}_{ij}-\mathcal{%
R}h_{ij})-h_{ij}\Box\Psi+\Psi_{;ij}\right)
\end{equation*}
where we denote $\Psi=\sqrt{\frac{2}{\mathcal{R}}}$. If the boundary
geometry has an isometry generated by a Killing vector $\xi ^{i}$, then $%
T_{ij}\xi ^{j}$ is divergence free, from which it follows that the quantity
\begin{equation*}
\mathcal{Q}=\oint_{\Sigma }d^{3}S^{i}T_{ij}\xi ^{j},
\end{equation*}%
associated with a closed surface $\Sigma $, is conserved. Physically, this
means that a collection of observers on the boundary with the induced metric
$h_{ij}$ measure the same value of $\mathcal{Q}$, provided the boundary has
an isometry generated by $\xi $. In particular, if $\xi ^{i}=\partial
/\partial t$ then $\mathcal{Q}$ is the conserved mass $\mathcal{M}$. One should also note that for KK black holes there exists another conserved quantity, analogous to the gravitational tension in the black string case \cite{Kastor:2006ti,Harmark:2004ch}, which can be easily computed in the counterterms approach by using the formula \cite{Stelea:2008tt}:
\begin{equation}
\mathcal{T'}=\int_{\Sigma'}d^{3}S_{\chi}T^{\chi}_{j}\xi ^{j}=\int dt\oint_{S^2_{\infty}}d^2x\sqrt{\sigma}T^{\chi}_{\chi},
\end{equation}
where now $\xi ^{i}=\partial/\partial \chi$ and the integration is performed over the two-sphere at infinity (described by $\theta$ and $\phi$) and also along the time direction. This gravitational tension is defined with respect to the asymptotic spatial translation along the circle described by $\chi$. Similarly to the black string case, one notices that one can omit the integration over time and work with the `tension per unit time':
\beqs
\mathcal{T}=\oint_{S^2_{\infty}}d^2x\sqrt{\sigma}T^{\chi}_{\chi}.
\eeqs

A straightforward computation using the boundary stress-tensor leads to the following conserved mass and gravitational tension:
\beqs
{\cal M}&=&\frac{\pi r_{\infty}}{G}\big[3(M_1+M_2)+\sqrt{r_{\infty}^2+(\sigma_1+\sigma_2)^2}\big],~~~{\cal T}=\frac{\sqrt{r_{\infty}^2+(\sigma_1+\sigma_2)^2}}{2G}.
\eeqs
The total charge of the double-black hole solution is computed using Gauss' formula with the result ${\cal Q}=\frac{\sqrt{3}\pi r_{\infty}(Q_1+Q_2)}{G}$.

Finally, let us notice that, similar to what happens in the single black hole case, one can compute the Komar mass:
\beqs
M_{K}&=&-\frac{1}{16\pi G }\frac{3}{2}\int_{S}\alpha,
\label{MK}
\eeqs
where $S$ is the boundary of any spacelike hypersurface and:
\begin{equation}
\alpha _{\mu \nu \rho }=\epsilon _{\mu \nu \rho \sigma \tau }\nabla ^{\sigma
}\xi ^{\tau }\ ,
\end{equation}%
with the Killing vector $\xi =\partial /\partial t$. This quantity is a measure of the mass contained in $S$, and if we take $S$ to be the three-sphere at infinity enclosing both horizons then (\ref{MK}) gives the total Komar mass of the system:
\beqs
M_K&=&\frac{3\pi (M_1+M_2)  r_{\infty}}{G}=\frac{3(M_1+M_2){\cal L}}{4G},
\eeqs
 while the Komar mass of each individual black hole is obtained by performing the above integration at the respective black hole horizon. At this point, let us note that the relation $2M_K=2{\cal M}-{\cal T}{\cal L}$ is satisfied for the double-black hole system as well (this relation was first noticed in \cite{Stelea:2008tt} for the single KK black hole system).

A straightforward computation leads to:
\begin{equation}
M_{Komar}^{(1)}=\frac{3}{4 G}\sigma _{2}{\cal L},~~~~~~~M_{Komar}^{(2)}=\frac{3}{4 G}\sigma _{1}{\cal L},
\end{equation}
while
\begin{equation}
M_K=M_{Komar}^{(1)}+M_{Komar}^{(2)}-\frac{1}{16\pi G}\frac{3}{2}\int R_{t}^{t}\sqrt{-g}dV.
\end{equation}
However, since Einstein's equations imply $R_{t}^{t}=\frac{F_{\mu t}^{2}}{3}$, one arrives at the following five-dimensional Smarr formula \cite{Gibbons:1987ps}:
\beqs
M_K&=&M_{K}^{(1)}+M_{K}^{(2)},
\eeqs
where for each constituent one has:
\beqs
2M_{K}^{(i)}=\frac{3k_{(5)}^{(i)}A^{(i)}_{(5)}}{8\pi G}+2\Phi_{H}^{(i)}{\cal Q}^{(i)}~,
\eeqs
where $M_{K}^{(i)}=\frac{3M_i{\cal L}}{4G}$. Thus one can regard $M_{K}^{(i)}$ as the individual mass of each black object, containing an electromagnetic contribution apart from the Komar piece! Here $k_{(5)}^i$, $A_{(5)}^i$, $\Phi_H^i$ and ${\cal Q}_i$ are in order the surface gravity, horizon area, electric potential and electric charge computed for each black hole horizon. One should also note that this relation follows from the four-dimensional Smarr relation for each individual black hole and in what follows, using the recent results obtained in \cite{Manko:2008gb} we shall show that this is indeed the case.

First, the electric potential on each horizon is easily computed from $\Phi_H^i=-A_t|_{horizon}$ and one obtains:
\beqs
\Phi_H^i&=&\sqrt{3}\left(\frac{M_i-\sigma_i}{Q_i}\right).
\eeqs
Also, the individual black hole charges turn out to be:
\beqs
{\cal Q}_i&=&\frac{\sqrt{3}\pi r_{\infty}Q_i}{G}=\frac{\sqrt{3}{\cal L}Q_i}{4G}.
\eeqs
Note that the total charge is ${\cal Q}={\cal Q}_1+{\cal Q}_2$, that is the sum of the individual charges, as expected.

The main difficulty when discussing the thermodynamic properties of the double black hole system consists in computing the temperature and entropy for each horizon. However, in what follows we shall show that these quantities are related in a simple manner to those defined in the original four-dimensional seed and, therefore, using the results in \cite{Manko:2008gb} one can express the five-dimensional ones in a simple form.

In the four-dimensional seed, the area of each black hole horizon can be expressed in the following form:
\beqs
A_{(4)}^i&=&4\pi\sigma_i\left(\rho\tilde{f}^{-1}e^{\tilde{\mu}}\right)|_{\rho=0}^i,
\eeqs
where for each black hole horizon we have \cite{Manko:2008gb}:
\beqs
\left(\rho\tilde{f}^{-1}e^{\tilde{\mu}}\right)|_{\rho=0}^1&=&\frac{\big[(R+M_1+M_2)(M_1+\sigma_1)-Q_1(Q_1+Q_2)\big]^2}{\sigma_1[(R+\sigma_1)^2-\sigma_2^2]},\nonumber\\
\left(\rho\tilde{f}^{-1}e^{\tilde{\mu}}\right)|_{\rho=0}^2&=&\frac{\big[(R+M_1+M_2)(M_2+\sigma_2)-Q_2(Q_1+Q_2)\big]^2}{\sigma_2[(R+\sigma_2)^2-\sigma_1^2]}.
\label{rhofmu}
\eeqs

For the final five-dimensional solution the area of each black hole horizon can be written as:
\beqs
A_{(5)}^i&=&4\pi \sigma_i{\cal L}\big[(\rho\tilde{f}^{-1}e^{\tilde{\mu}})|_{\rho=0}^i\big]^{\frac{3}{4}}\left((\rho^{\frac{1}{2}}e^{2\gamma-2h}\Sigma)|_{\rho=0}^i\right)^{\frac{1}{2}}.
\label{area}
\eeqs
Let us note now that near each black hole horizon one can express:
\beqs
e^{2h-2\gamma}&=&(p_i)^2\sqrt{\rho}+{\cal O}(\rho)
\eeqs
where $p_i$ are constants, while $\Sigma|_{\rho=0}=\Sigma_0=1+c^2$ is also constant. One finds explicitly:
\beqs
(p_1)^2&=&\frac{1}{2}\sqrt{\frac{R+\sigma_1-\sigma_2}{\sigma_1(R+\sigma_1+\sigma_2)}},~~~(p_2)^2=\frac{1}{2}\sqrt{\frac{R+\sigma_2-\sigma_1}{\sigma_2(R+\sigma_1+\sigma_2)}}
\eeqs
and replacing these relations in (\ref{area}) one finally finds:
\beqs
A_{(5)}^1&=&4\pi\sigma_1{\cal L}\left(1+\frac{\sqrt{r_{\infty}^2+(\sigma_1+\sigma_2)^2}}{\sigma_1+\sigma_2}\right)^{\frac{1}{2}}\bigg[\frac{\big[(R+M_1+M_2)(M_1+\sigma_1)-Q_1(Q_1+Q_2)\big]^3}{\sigma_1(R+\sigma_1+\sigma_2)(R+\sigma_1-\sigma_2)^2}\bigg]^{\frac{1}{2}},\nonumber\\
A_{(5)}^2&=&4\pi\sigma_2{\cal L}\left(1+\frac{\sqrt{r_{\infty}^2+(\sigma_1+\sigma_2)^2}}{\sigma_1+\sigma_2}\right)^{\frac{1}{2}}\bigg[\frac{\big[(R+M_1+M_2)(M_2+\sigma_2)-Q_2(Q_1+Q_2)\big]^3}{\sigma_2(R+\sigma_1+\sigma_2)(R+\sigma_2-\sigma_1)^2}\bigg]^{\frac{1}{2}}\nonumber
\eeqs

As a check of the correctness of the above formulae for the horizon areas, let us take for instance $M_1=Q_1=0$ and further send $R\ra\infty$. One readily checks that one obtains the horizon area of the remaining black hole:
\beqs
A^2_{(5)}&=&4\pi{\cal L}(M_2+\sigma_2)^{\frac{3}{2}}\sqrt{\sigma_2+\sqrt{r_{\infty}^2+\sigma_2^2}}
\eeqs
as expected.

In order to compute the Hawking temperature we shall use the definition in terms of the surface gravity, which is generally defined as $k^2=-\frac{1}{2}\xi^{a;b}\xi_{a;b}$, where $\xi=\partial/\partial t$ is the canonically normalized timelike Killing vector. From the general expressions in (\ref{Manko}) one can deduce that near each black hole horizon one has the following expansions:
\beqs
\tilde{f}=F(z)\rho^2+{\cal O}(\rho^3),~~~~e^{2\tilde{\mu}}=X(z)\tilde{f}+{\cal O}(\rho^3).
\eeqs
Replacing these expressions into the four-dimensional\footnote{This is computed using the seed metric in four dimensions.} surface gravity and taking the $\rho\ra 0$ limit one finds the particularly simple result:
\beqs
k_{(4)}&=&\sqrt{\frac{F(z)}{X(z)}}.
\eeqs
Let us note at this point that the above expression for the surface gravity is actually independent of $z$ as expected (since it should be constant on the black hole horizon) and one can confirm this by computing:
\beqs
(\rho\tilde{f}^{-1}e^{\tilde{\mu}})|_{\rho=0}&=&\sqrt{\frac{X(z)}{F(z)}}=\frac{1}{k_{(4)}},
\eeqs
which is manifestly constant according to (\ref{rhofmu}). Then the Hawking temperature is $T_{(4)}^i=\frac{k_{(4)}^i}{2\pi}$ and one obtains:
\beqs
\frac{k_{(4)}^iA_{(4)}^i}{8\pi G_{(4)}}=\frac{\sigma_i}{2G_{(4)}},
\eeqs
where $G_{(4)}=G/{\cal L}$ is Newton's constant in four dimensions.

If one computes the surface gravity for the five-dimensional solution one obtains the simple form:
\beqs
k_{(5)}^i&=&\frac{p_i}{\sqrt{\Sigma_0}}(k_{(4)}^i)^{\frac{3}{4}}.
\eeqs
It is now an easy matter to check that:
\beqs
\frac{A_{(5)}^ik_{(5)}^i}{8\pi G}&=&\frac{{\cal L}\sigma_i}{2G}=\frac{\sigma_i}{2G_{(4)}}=\frac{k_{(4)}^iA_{(4)}^i}{8\pi G_{(4)}}.
\eeqs

It is now apparent that the individual Smarr relations for each black hole:
\beqs
2M_{K}^i=3\left(\frac{A_{(5)}^ik_{(5)}^i}{8\pi G}\right)+2\Phi^i{\cal Q}_i
\eeqs
are satisfied and moreover they are equivalent to the Smarr relations in the initial four-dimensional seed as advertised.

Finally, making use of the relationship between the total Komar mass and the mass computed in the counterterm approach, one can rewrite the Smarr relation for the double KK black hole as:
\beqs
2{\cal M}&=&3\left(\frac{A_{(5)}^1k_{(5)}^1}{8\pi G}+\frac{A_{(5)}^2k_{(5)}^2}{8\pi G}\right)+{\cal T}{\cal L}+2(\Phi^1{\cal Q}_1+\Phi^2{\cal Q}_2)
\eeqs

For a single KK black hole this Smarr relation reduces to the one previously obtained in \cite{Stelea:2008tt} as expected.

\section{Conclusions}

By using a simple modification of novel solution generation technique described in \cite{Chng:2006gh}, we were able to construct the general non-extremally charged KK multi-black hole solutions in five dimensions. In particular, this new technique provides us with a mapping between static charged four-dimensional solutions of the Einstein-Maxwell system to five-dimensional charged and static Einstein-Maxwell-dilaton solutions with KK asymptotics. While the general solution of the EMD system can be read in (\ref{final5dalpha}), in this paper, when discussing the generated solutions we focused for simplicity on Einstein-Maxwell theory, for which the coupling constant $\alpha=0$ in the general solution (\ref{final5dalpha}) vanishes.

In Section $2$ we described this solution generating technique, while in Section $3$ we generated KK multi-black hole systems in five dimensions. In the single KK black hole case, it turns out that, instead of using the charged black hole with spherical symmetry as the initial seed, in order to obtain sensible results one has to use the charged single black string solution. Motivated by this result, we then used as seed the general double black string solution in an asymptotically flat background in order to generate the general charged double-black hole solution in the KK background. We investigated the conical singularity structure of the final solution and we also showed how to recover several known solutions as particular cases. In particular we showed that the unconstrained limit yields 2 extremal black string solutions but the alternate limit, in which one takes the limit with the constraints that $ 2c^2 \sigma_1\equiv N_1$ and $2 c^2 \sigma_2 \equiv N_2$ are fixed yields another, the extremal double black hole solution. Finally, we computed its conserved charges at infinity and discussed at length its thermodynamic properties. In particular, based on the previous results recently derived in \cite{Manko:2008gb} for the initial four-dimensional seed solution,  we proved the general Smarr relation for the double black hole system in five dimensions.

As avenues for further work, it would be interesting to identify the seed solution which will lead to the construction of a black ring system in the KK-background. This last solution has been recently constructed in \cite{Camps:2008hb} and one should be able to recover it using the methods presented in this paper. 

Another interesting possibility is to extend the analysis of \cite{Costa:2000kf} to five dimensions and find an embedding of the KK multi-black hole solution in string theory. Using the effective string description one should be able to compute for instance the entropy, including the corrections associated with the interaction among the black holes.

\vspace{10pt}

{\Large Acknowledgements}

This work was supported by the Natural Sciences and Engineering Council of Canada. CS would like to thank Eugen Radu for interesting remarks on the manuscript.

\end{document}